\documentclass[sigconf, natbib]{acmart}

\usepackage[utf8]{inputenc}
\usepackage{url}
\usepackage{listings}
\usepackage{subcaption}
\usepackage{bbding}
\usepackage{pifont}
\usepackage{array}
\usepackage{siunitx}

\usepackage{caption}
 
\usepackage{newfloat}
\DeclareFloatingEnvironment[
   fileext=lst,
   name=Listing
]{listing}
\floatname{listing}{Listing}

\newcommand\code[1]{{\small{\texttt{#1}}}}

\usepackage{xspace}
\usepackage{etoolbox} % for \robustify
\newcommand{\CC}{C\nolinebreak\hspace{-.05em}\raisebox{.2ex}{\small +}\nolinebreak\hspace{-.05em}\raisebox{.2ex}{\small +}\xspace}

\robustify{\CC}
\newcommand{\ompi}{Open~MPI\xspace}

\title{Quo Vadis MPI RMA? Towards a More Efficient Use of MPI One-Sided Communication}
\acmYear{2021}
\acmConference[EuroMPI'21]{EuroMPI 2021}{September 7th, 2021}{Garching, Germany}

\author{Joseph Schuchart}
\email{schuchart@icl.utk.edu}
\additionalaffiliation{
  \institution{High-Performance Computing Center Stuttgart (HLRS)}
  \streetaddress{Nobelstraße 19}
  \postcode{70597}
  \city{Stuttgart}
  \country{Germany}
}
\affiliation{
  \institution{Innovative Computing Laboratory (ICL), University of Tennessee Knoxville (UTK)}
  \streetaddress{1122 Volunteer Blvd}
  \city{Knoxville}
  \state{TN}
  \postcode{37996}
  \country{U.S.A.}
}
\author{Christoph Niethammer}
\email{niethammer@hlrs.de}
\affiliation{
  \institution{High-Performance Computing Center Stuttgart (HLRS)}
  \streetaddress{Nobelstraße 19}
  \postcode{70597}
  \city{Stuttgart}
  \country{Germany}
}

\author{Jos\'{e} Gracia}
\email{gracia@hlrs.de}
\affiliation{
  \institution{High-Performance Computing Center Stuttgart (HLRS)}
  \streetaddress{Nobelstraße 19}
  \postcode{70597}
  \city{Stuttgart}
  \country{Germany}
}

\author{George Bosilca}
\email{bosilca@icl.utk.edu}
\affiliation{
  \institution{Innovative Computing Laboratory (ICL), University of Tennessee Knoxville (UTK)}
  \streetaddress{1122 Volunteer Blvd}
  \city{Knoxville}
  \state{TN}
  \postcode{37996}
  \country{U.S.A.}
}

\keywords{MPI-RMA, Memory Handles, MPI, RDMA}

\graphicspath{{figures/}}

\begin{document}

\begin{abstract}
The MPI standard has long included one-sided communication abstractions through the MPI Remote Memory Access (RMA) interface.
Unfortunately, the MPI RMA chapter in the 4.0 version of the MPI standard still contains both well-known and lesser known short-comings for both implementations and users, which lead to potentially non-optimal usage patterns.
In this paper, we identify a set of issues and propose ways for applications to better express anticipated usage of RMA routines, allowing the MPI implementation to better adapt to the application's needs.
In order to increase the flexibility of the RMA interface, we add the capability to duplicate windows, allowing access to the same resources encapsulated by a window using different configurations.
In the same vein, we introduce the concept of MPI memory handles, meant to provide life-time guarantees on memory attached to dynamic windows, removing the overhead currently present in using dynamically exposed memory.
We will show that our extensions provide improved accumulate latencies, reduced overheads for multi-threaded flushes, and allow for zero overhead dynamic memory window usage.
\end{abstract}
\maketitle

\section{Introduction}
\label{sec:intro}

Modern high-performance networks commonly provide the capability to directly access memory on a remote host for reading, writing, and atomic memory updates~\cite{Alverson:2012:cray,ConnectIB}.
The hardware is capable of transferring data without the involvement of the CPU at the target node once the upper software layers have properly set up the parameters for the transfer, e.g., registered memory with the network interface card (NIC) and exchanged the registration information with the peers involved.
MPI implementations typically make use of these low-level network features to provide efficient transfer of large messages between peers communicating through point-to-point or collective operations~\cite{Sur:2006:RRBR}.

The MPI RMA interface was introduced with MPI version 2.0~\cite{mpi2.0} and has seen a major overhaul in version 3.0~\cite{mpi3.0}, including the addition of allocated and dynamic windows.
The intention of this interface is to expose the network's low-level remote direct memory access (RDMA) capabilities to the user by providing procedures for \emph{put}, \emph{get}, and \emph{accumulate} operations on \emph{windows} that encapsulate memory for which registration information has been exchanged among the group of participating peers.
By using MPI RMA, applications are able to decouple communication and synchronization, e.g., to perform bursts of communication before synchronizing through collective and point-to-point communication or by setting a signal flag at the target using accumulate operations.

In its current form, an MPI window is an object spanning across the processes in its group, i.e., its creation and destruction are collective operations.
Active target synchronization involves collective operations.
Passive target synchronization, on the other hand, only involves specific MPI calls at the origin of the operation (some MPI implementation may, however, depend on the target to call into MPI procedure to progress outstanding RMA operation).
With the exception of dynamic windows, the memory accessed through these windows is static, requiring collective (re)allocations to increase the amount of memory accessible to RMA operations.
Dynamic windows, on the other hand, allow for dynamically attaching and detaching memory segments, albeit at a significant penalty in performance, which will be discussed in \autoref{sec:state}.

Interest in RMA in the user community seems to be growing~\cite{Berholdt:2018:MPIUsage,Daliss:2019:FPD}.
However, the RMA chapter has seen little change during the work on the 4.0 release of the standard, despite there being several known shortcomings that inhibit full and efficient usage of MPI RMA in applications and runtime systems~\cite{Schuchart:2018:REU, Schuchart:2019:UMR}.
In this paper, we draw from our experience in using MPI-3 RMA in the context of the DASH project~\cite{Fuerlinger:2016:DASH, Schuchart:2018:REU} in general and the global task synchronization scheme built on top of it in particular~\cite{Schuchart:2019:GTD}.
We will discuss a number of issues found during this work and propose potential remedies that mostly consist in allowing the user to express the anticipated use of the RMA interface to the implementation through additional info keys (\autoref{sec:user_info}).
In order to increase flexibility in applying these configurations, we propose an extension to the MPI RMA interface that allows users to duplicate windows, accessing the same memory using the same network resources but with configurations adjusted to the needs of different regions of the application (\autoref{sec:window_dup}).
In order to increase flexibility in the use of RDMA through the MPI RMA interface, we propose the addition of \emph{MPI memory handles} that allow users to explicitly manage the registration information of memory attached to dynamic windows, alleviating the performance penalties that stem from the current design of dynamic windows (\autoref{sec:dmh}).

A brief discussion of additional improvements and changes that are beyond the scope of this work will be provided in \autoref{sec:additional_improvements}.
\autoref{sec:implementation} discusses some of the implementation details for the proposed solutions before an evaluation of some of the proposals using micro benchmarks %and an integration into the PaRSEC data-flow engine~\cite{Bosilca:2011:FDD}
is presented in \autoref{sec:evaluation}.
Related work is discussed in \autoref{sec:related_world} and conclusions are drawn in \autoref{sec:conclusions}.

\section{Additional User-Provided Information}
\label{sec:user_info}
\subsection{Thread-Scope Synchronization}
\label{sec:thread_local_flush}

Communication and synchronization in MPI RMA happens at the scope of processes, which encapsulate the memory made accessible to remote peers.
Thus, a flush in passive target synchronization or a fence in active target synchronization ensure the completion of all operations previously issued by any thread in the current process to complete in the target process memory.
Active target synchronization is collective either over the group of a window (\code{MPI\_Win\_\-fence}) or an otherwise provided group (post-start-complete-wait, PSCW).
With passive target synchronization, flushes are local operations.
Despite previous attempts at thread-specific communication endpoints~\cite{Dinan:2013:EMI}, collective operations in MPI happen at the scope of processes.
We will thus focus on the passive target synchronization.

\begin{figure}
\begin{subfigure}{.45\columnwidth}
\includegraphics[width=\textwidth, page=1]{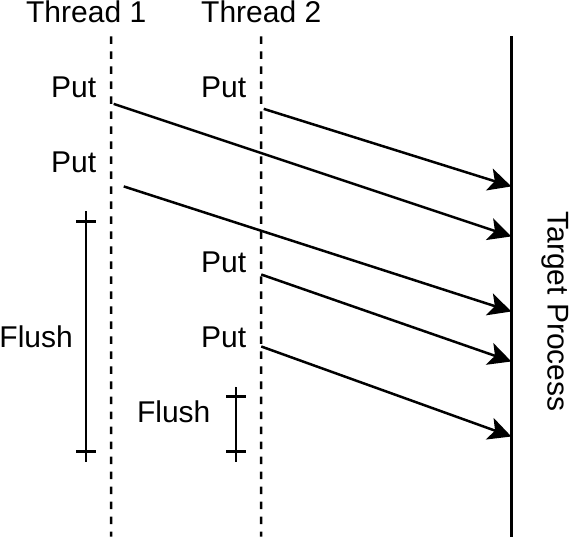}
\caption{Process-scope.}
\label{fig:flush_scope:process}
\end{subfigure}
\hfill
\begin{subfigure}{.45\columnwidth}
\includegraphics[width=\textwidth, page=2]{figures/thread_process_scope-crop.pdf}
\caption{Thread-scope.}
\label{fig:flush_scope:thread}
\end{subfigure}
\caption{Flushes with process- and thread-scope. With process-scope flushes, Thread 1 potentially waits for the completion of all of its operations and the operations issued by Thread 2.}
\label{fig:flush_scope}
\end{figure}

In multi-threaded applications, individual threads may perform RMA operations independently.
However, a thread calling into \code{MPI\_Win\_flush} potentially waits for the completion of operations issued by all threads of the same process to the same target, even though the completion of operations previously issued by the current thread might be sufficient for the application.
This is depicted in \autoref{fig:flush_scope:process}, where the flush of Thread 1 is prolonged by operations issued by Thread 2.
MPI implementations may use thread-specific network hardware resources (\emph{endpoints} or \emph{rails}) to reduce synchronization between threads when issuing RMA operations, e.g., by using one endpoint per thread or distributing operations across a fixed number of endpoints~\cite{Hjelm:2018:IMM}.

MPI supports flushing operations to a single target or to all targets in the group of the window, either with local or remote completion.
We will focus on operations with remote completion, although the proposal also applies to flushes with local completion.

In order to restrict the completion semantics of flush operations to operations previously issued by the calling thread, the user has to signal this intention to the implementation.
While it would be feasible to introduce a new set of functions to accomplish this goal, the required extension of the API would force implementations to provide such functionality even if support for multi-threaded RMA was limited.
It would also introduce the notion of thread-scope operations into an API that is otherwise oblivious of the existence of multiple threads of execution, with the exception of \code{MPI\_Init\_thread} used to signal their (anticipated) existence.

We thus propose the addition of an info key called \code{mpi\_win\_scope}, which specifies the scope of synchronization operations on a window.
If the value of that key is set to \code{process} (the default), flushes behave as today with operations issued by all threads required to complete during a flush.
However, if the scope is set to \code{thread}, the implementation is free to restrict the scope of a flush to the operations previously issued by the calling thread, as depicted in \autoref{fig:flush_scope:thread}.
Since the \code{process} scope is a superset of the \code{thread} scope, implementations ignoring this info key remain correct.
Users can check the support for the \code{thread} scope by querying the value associated with that key from the info object attached to the window using \code{MPI\_Win\_get\_info}~\cite[\S12.2.7]{mpi4.0}.

With the \code{mpi\_win\_scope} key set to \code{thread} on a window, implementations using thread-local endpoints only need to wait for the completion of operations on the endpoint assigned to the calling thread, potentially avoiding any synchronization between threads using RMA with passive target synchronization.
The key has no effect on active target synchronization, since collective operations always happen at the process scope.

\subsection{Operation Ordering}
\label{sec:oporder}

By default, MPI RMA guarantees the ordering of consecutive accumulate operations on the same memory location with the same data type and allows users to relax this constraint using the \code{accumu\-late\_ordering} info key.
However, in order to achieve ordering of put and get operations or to order accumulate operations to distinct memory location within a window, the application is required to wait for all outstanding operations to complete before issuing operations that are required to occur later in the sequence.
This completion might entail at least the latency of a full round-trip in the network, depending on the number of previously issued operations.

\begin{listing}
\begin{lstlisting}
int flag, one = 1;
MPI_Request req;
MPI_Rput(..., target, win, &req);
do {
  do_useful_work();
  MPI_Test(&req, &flag, MPI_STATUS_IGNORE); |\label{line:put_inc:test1}|
} while (!flag);
/* Flush needed for remote completion */
MPI_Win_flush(target, win);       |\label{line:put_inc:flush}|
/* Signal that the data has been written by
 * incrementing a counter at the target */
MPI_Raccumulate(&one, ..., target, ..., win, &req);
do {
  do_useful_work();
  MPI_Test(&req, &flag, MPI_STATUS_IGNORE);  |\label{line:put_inc:test2}|
} while (!flag);
\end{lstlisting}
\caption{Using an atomic increment to signal the completion of a put, overlapping communication with useful work.}
\label{lst:put_inc}
\end{listing}

\autoref{lst:put_inc} provides an example in which an accumulate operation is used to signal the availability of data previously put into the target's memory.
In order to hide the latency of both operations, the application tests on requests for both operations (Lines~\ref{line:put_inc:test1} and ~\ref{line:put_inc:test2}) and requires a flush in between (Line~\ref{line:put_inc:flush}) to ensure remote completion before the signal is set.

Modern high-performance networks provide so-called \emph{fence} operations, allowing users to request the hardware to order the completion of two operations $Op_1$ and $Op_2$ in the order in which they were issued, similar to a memory barrier in shared memory systems.
We have previously proposed adding a function \code{MPI\_Win\_order}~\cite{Schuchart:2019:UMR}, which would translate either into a memory barrier or a fence in the network interface card.
In multi-threaded applications, however, the default process-scope of MPI RMA would require a fallback to waiting for completion of all operations if thread-specific network resources are used.
Using the thread-local scope proposed in \autoref{sec:thread_local_flush} may provide a partial solution by constraining the scope of ordering to operations issued by individual threads.
However, in cases where the implementation issues operations of a single thread to multiple hardware resources (e.g., for explicit load-balancing of communication) multiple streams of operations would again have to be synchronized by waiting for completion of prior operations.

The underlying problem, however, is that the ordering request may be injected into the operation stream $Op_n, \ldots, O_{m}$  at any time.
As a consequence, the MPI implementation has no prior knowledge of the ordering request at the time $Op_n$ is issued and thus would have to either constrain itself to using configurations in which operations can safely be ordered or resort to waiting for completion to enforce ordering.

In order to provide the MPI implementation with \emph{a priori} information about intended operation ordering we propose the concept of \emph{ordered operation sequences}.
The user thereby signals the request to order a set of operations \emph{before} issuing the first operation included in that sequence.
We propose a new info key that enables operation ordering on a given window, called \code{mpi\_win\_order}.
While set to \code{true}, the sequence of operations issued to the same target on this window will complete at the target in the order in which the operations were issued.
This provides sufficient information to the implementation to implement operation ordering without resorting to completion in the middle of the sequence, albeit potentially at the cost of using a single endpoint.
However, in combination with the \code{thread} scope described previously ordering can be constrained to operations issued by the same thread.

Applications relying on ordering using this info key are required to check whether the operation ordering using the \code{mpi\_win\_order} info key is supported and fall back to flushes in between operations otherwise.
\autoref{lst:put_inc_order} provides a modified version of the example in \autoref{lst:put_inc} with the \code{mpi\_win\_order} set to \code{true}, avoiding the intermittent flush and only testing for the accumulate operation request.

\begin{listing}
\begin{lstlisting}
int flag, one = 1;
MPI_Request req;
MPI_Put(..., target, win);
/* Signal that the data has been written by
 * incrementing a counter at the target */
MPI_Raccumulate(&one, ..., target, ..., win, &req);
do {
  do_useful_work();
  MPI_Test(&req, &flag, MPI_STATUS_IGNORE);
} while (!flag);
\end{lstlisting}
\caption{The example of \autoref{lst:put_inc} with \code{mpi\_win\_order} set to \code{true}, avoiding the first flush by chaining two operations.}
\label{lst:put_inc_order}
\end{listing}

\subsection{Hardware Accumulate Operations}
\label{sec:hardware_accumulate}

MPI RMA accumulate operations are notoriously hard to implement efficiently: on the one hand, single-element operations such as \code{MPI\_Fetch\_and\_op} and \code{MPI\_Compare\_and\_swap} may benefit from the use of hardware atomic operations provided by the NIC due to the low latency of single-element operations implemented by the hardware.
On the other hand, \code{MPI\_Accumulate} allows users to issue operations on an arbitrary number of elements at the same time, which could potentially benefit from the higher bandwidth of operations performed by vector units on the host CPU~\cite{Zhong:2020:AVX}.
The MPI standard requires implementations to provide element-wise atomicity of operations applied to the same memory location if using the same data type, regardless of whether they were issued through single-element operations or as part of a multi-element operation.
In addition, not all possible operations are supported by the network hardware.
For example, while networks commonly support addition and subtraction of integral values, support for integral value multiplication or operations on floating point values is often missing.
Taken together, implementations once again cannot anticipate the operations that will be issued by the user and thus have to leave certain hardware features lay bare.

Implementation typically use two possible approaches: i) taking a lock at the target process before fetching the data, applying the operation, and writing the data back before releasing the lock; and ii) using active messages to transfer the data to the target and relying on the target CPU to perform the operation.
Both approaches require the serialization of concurrent accumulate operations through some form of mutual exclusion device.

An existing proposal to tackle this problem allows the user to specify how many elements will be used with which operations~\cite{Si:2018:RMAInfo}.
However, users would still have no information on whether accumulate operations will be executed in hardware or software.
Some applications may rely on low-latency accumulate operations~\cite{Schuchart:2019:UMR, Brock:2019:BCL} but the lack of transparency prevents them from picking an alternative algorithm if available.
The proposed unidirectional signaling is hence not sufficient.

We propose the addition of a procedure (based on a previous suggestion in the MPI RMA working group~\cite{Dinan:2018:Query}) that i) allows the application to query the implementation's approach to performing a given accumulate operations for a given number of elements of a certain data type; and ii) to signal the anticipated usage pattern to the implementation.
We borrow a concept from the \CC{} \code{std::atomic} wrapper type~\cite{ISO:14882:2011}, which allows developers to query whether atomic modification of a given wrapped type is \emph{lock-free} using the compile-time \code{is\_lock\_free} trait that signals whether a mutex or CPU-provided atomic operations are used to ensure atomicity.
Contrary to \code{std::atomic}, MPI accumulate operations may apply to multiple elements at once and implementations may use a threshold for switching between hardware and software approaches.

\begin{listing}
\begin{lstlisting}
/**
 * Query whether the implementation employs hardware operations
 * intrinsic to the origin node to perform the operations listed
 * in ops on a maximum of max_count elements of type on the provided
 * window.
 * Flag will be set to 1 if intrinsic hardware operations at the
 * origin are used to perform these operations and 0 otherwise.
 */
int MPI_Win_op_intrinsic(const char *ops, MPI_Aint max_count,
                         MPI_Datatype type, MPI_Win win, int *flag);
\end{lstlisting}
\caption{Function to query the use of intrinsic hardware operations for a given set of operations on a number of elements of a certain data type.}
\label{lst:lockfree}
\end{listing}

Thus, the information has to be query-able at runtime, for which we propose a new procedure called \code{MPI\_Win\_op\_intrinsic} listed in \autoref{lst:lockfree}.
For a given tuple describing the set of anticipated operations to be performed (\code{ops}) on the provided number elements (\code{max\_count}) of a certain \code{type} on a specific window \code{win}, the implementation returns whether the operations will be performed with hardware operations intrinsic to the origin node, i.e., without relying on the participation of a CPU at the target.\footnote{We note that an accumulate operation using the network hardware technically relies on a processor at the target to perform the operation. However, the accumulate instruction is issued to the NIC at the origin and is thus intrinsic to the origin hardware.}
The set of operations are described as a string containing a comma-delimited list of operations, using the second half of the name of predefined \code{MPI\_Op} elements (e.g., ``sum''), ``replace'' for \code{MPI\_REPLACE}, and ``cas'' to denote \code{MPI\_Compare\_and\_swap}.

The information obtained from a call to \code{MPI\_Win\_op\_intrinsic} may then be used to set a new boolean info key called \code{mpi\_assert\_\-accu\-mulate\_intrinsic}.
If set to true, the application asserts that it will only issue accumulate operations in configurations for which the implementations has signaled the use of intrinsic operations.
The results of the application disregarding this assertion are undefined, leading to modifications that are potentially non-atomic.

With this bidirectional signaling mechanism we achieve two goals: a) providing transparency to the application on how the MPI implementation will handle a given configuration of accumulate operations; and b) allowing the application to announce their anticipated behavior to the implementation.
This, in turn, will allow implementations to safely make use of hardware atomic operations if all anticipated operations used by the applications can be mapped onto the hardware and the number of elements is below the threshold controlling the switch in the trade-off between low latency and high bandwidth.

\section{Duplicating MPI Windows}
\label{sec:window_dup}

In the previous section we have proposed three new info keys to help the user express the intended use of RMA operations on a window: restricting the scope of flushes to operations issued by the calling thread; ordering the operations issued on the window; and limiting the number of elements in accumulate operations to allow for the use of operations intrinsic to the origin hardware. %; and requesting that the completion of requests returned by \code{MPI\_Rput} entails remote completion.
However, changing the value of an info key overwrites the old value and thus makes it impossible to use different configurations concurrently, e.g., to request operation ordering on some threads but not on others or to use bandwidth-optimized accumulate operations on one part of the memory while using latency-optimized accumulate operations on another part of the window.
Switching between these settings would  require careful orchestration of info key values.

\begin{figure}
\begin{subfigure}{.48\columnwidth}
\centering
\includegraphics[width=.9\textwidth, page=1]{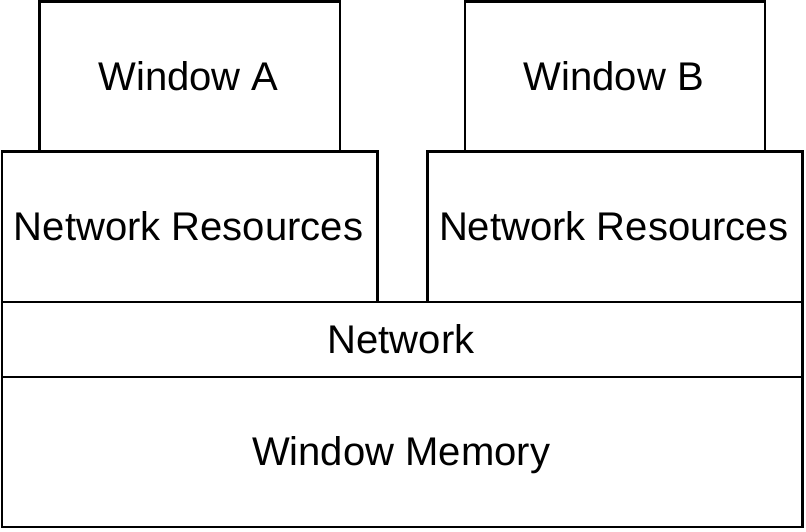}
\caption{Independent windows.}
\label{fig:window_dup_arch:independent}
\end{subfigure}
\hfill
\begin{subfigure}{.48\columnwidth}
\centering
\includegraphics[width=.8\textwidth, page=2]{figures/mpi-win-dup-crop.pdf}
\caption{Duplicated windows.}
\label{fig:window_dup_arch:dup}
\end{subfigure}
\caption{Two windows accessing the same window memory through the network.}
\label{fig:window_dup_arch}
\end{figure}

While it is legitimate to allocate memory in \code{MPI\_Win\_allocate} and pass that memory into a call to \code{MPI\_Win\_create} with different info key values, the resulting two windows are semantically independent with independent passive and active target synchronization semantics and no cross-window atomicity guarantees, as depicted in \autoref{fig:window_dup_arch:independent}.
In order to ease the task of managing different means of access to the same window memory and to keep windows with different info values  in sync, we propose to add a \emph{window duplication} function, as outlined in \autoref{lst:win_dup}.
In contrast to two independently created windows, duplicated windows may share internal data structures and window memory while carrying potentially different access semantics.
Duplicated windows can thus be considered as different handles to the same underlying memory and network resources, as depicted in \autoref{fig:window_dup_arch:dup}.

This approach enables the use case described above in which different window settings are used in different parts of an application, while sharing the underlying resources.
Some restrictions apply: as long as the duplicated windows use the same value for \code{mpi\_assert\_accu\-mulate\_intrinsic} accumulate operations are atomic across these windows.
However, issuing accumulate operations on two windows having different values for the \code{mpi\_assert\_\-accu\-mu\-late\_intrinsic} is legal but the accumulate operations may not be atomic with respect to each other.
It is up to the user to coordinate the correct use of this info key.

\begin{listing}
\begin{lstlisting}
/* Duplicate a window with different info key values.
 * Both the parent window and the new window access the same 
 * memory regions with potentially different configurations. */
int MPIX_Win_dup_with_info(MPI_Win  parentwin, 
                           MPI_Info info,
                           MPI_Win *newwin);
\end{lstlisting}
\caption{Signature of window duplication function.}
\label{lst:win_dup}
\end{listing}

We propose a new function called \code{MPIX\_Win\_dup\_with\_info} that is used to duplicate the window with a new set of info keys.
Its signature is shown in \autoref{lst:win_dup}.
Info keys from the parent window will be duplicated into the new window, with the provided info keys overriding existing ones.

Since the original and duplicated windows are not logically separate, all synchronization operations applied to a window also apply to its duplicates, and vice versa.
For example, the duplicated window may not be locked if the parent window has already been locked.
In essence, window duplication is akin to assignment of an \code{MPI\_Win} variable with the added ability to control certain info values.

The call to \code{MPIX\_Win\_dup\_with\_info} is a local operation and thus does not entail any synchronization with other processes in the group of the parent window.
An MPI implementation may not be able to change certain info keys during this call and may thus reject the change by retaining the original or default value.
Users should check whether the MPI implementation is able to support the requested configuration by querying the active info keys using \code{MPI\_Win\_get\_info}.

\section{Dynamic Memory Handles}
\label{sec:dmh}

\begin{figure}
\begin{subfigure}{.31\columnwidth}
\centering
\includegraphics[width=.9\textwidth, page=1]{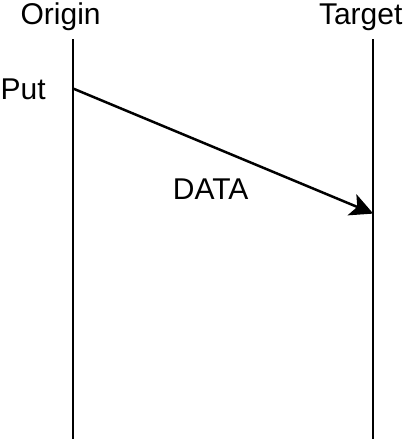}
\caption{Static window (RDMA)}
\label{fig:dynamic_win_put:rdma}
\end{subfigure}
\hfill
\begin{subfigure}{.31\columnwidth}
\centering
\includegraphics[width=.9\textwidth, page=2]{figures/dynamic_win_put-crop.pdf}
\caption{Dynamic window (RDMA)}
\label{fig:dynamic_win_put:fetch}
\end{subfigure}
\hfill
\begin{subfigure}{.31\columnwidth}
\centering
\includegraphics[width=\textwidth, page=3]{figures/dynamic_win_put-crop.pdf}
\caption{Dynamic window (active messages)}
\label{fig:dynamic_win_put:am}
\end{subfigure}
\caption{Possible implementations of put on RDMA-capable networks for static and dynamic windows.}
\label{fig:dynamic_win_put}
\end{figure}

In its current form, window creation is a collective operation in MPI.
With the exception of dynamic windows, MPI windows and their memory are statically allocated or assigned, which allows the MPI implementation to exchange all relevant connection and registration information during window creation and enables the use of the network's RDMA capabilities (\autoref{fig:dynamic_win_put:rdma}).
However, such static windows may be impractical if an application's communication requirements changes over time, requiring repeated (collective) reallocation of windows.
Moreover, applications may treat communication and memory allocation as orthogonal concerns such that the use of window memory would break through abstraction boundaries, requiring major restructuring efforts to integrate the allocation of memory through MPI windows.

In contrast, dynamic windows, as their name suggests, allow users to attach and detach memory dynamically in a local operation after the window has been created in a collective operation.
Attaching the memory explicitly allows the MPI implementation to register the memory with the network device for later access using RDMA.
Addressing in dynamic windows is done using absolute virtual addresses: after attaching memory to the dynamic window, the process distributes the virtual address that is then used as displacement at the origin of an RMA operation.

Unfortunately, the use of virtual addresses is the greatest weakness of dynamic windows: in contrast to static windows, the origin of an RMA operation initially has no information on the underlying memory registration and thus has to either query this information from the target before issuing RDMA operations (\autoref{fig:dynamic_win_put:fetch}) or fall-back to emulating remote memory accesses using active messages (AM, \autoref{fig:dynamic_win_put:am}).
Both approaches add considerable latency, especially to RMA operations on small amounts of data.
Since the target may detach and reattach the same virtual base address, the registration information for the same virtual address may change at the target in between RMA operations at the origin.
Thus, while caching techniques are possible, the origin has to at least verify the validity of the cached registration information on \emph{every} RMA communication operation.
The lack of \emph{life-time guarantees} is thus the main reason for added overhead when using dynamic windows.

We will show in \autoref{sec:state} that the difference in latency between allocated and dynamic windows is significant on all tested MPI implementations.
We will then propose an extension to the MPI RMA interface to allow applications to explicitly exchange registration information and thus make life-time guarantees to the MPI implementation that enable it to use RDMA with zero overhead.

\subsection{State of the Art}
\label{sec:state}

\autoref{tab:software} lists the software used for comparison of dynamic and allocated windows in this section.
All measurements were conducted on an HPE Apollo 6500 system \emph{Hawk} installed at HLRS.\footnote{More details at \url{https://www.hlrs.de/systems/hpe-apollo-hawk/}}
The nodes are equipped with dual-socket 64-core AMD EPYC 7742 processors and connected through Mellanox InfiniBand HDR200 in a 9D hyper-cube fabric.

\begin{table}
\renewcommand{\arraystretch}{1.1}
\caption{Software configuration.}
\label{tab:software}
\begin{tabular}{llp{.45\columnwidth}}
\toprule
\textbf{Software} & \textbf{Version} & \textbf{Configuration/Remarks} \\
\midrule
\ompi & \code{4.0.5} & \code{--with-ucx=...} \\
MPICH & \code{v4.0.a} & \code{--with-device=ch4:ucx}  \\
MVAPICH & \code{2.3.5} & \code{--with-device=ch3:mrail --with-rdma=gen2} \\
UCX & 1.10.0 & \code{--enable-mt --with-xpmem --with-verbs --with-rdmacm} \\
GCC & 10.2.0 & \textit{site installation} \\
OSU Benchmarks & 5.6.2 & \emph{none} \\
\bottomrule
\end{tabular}
\end{table}

\subsubsection{Communication Latency}

\begin{figure}
\centering
\includegraphics[width=.9\columnwidth]{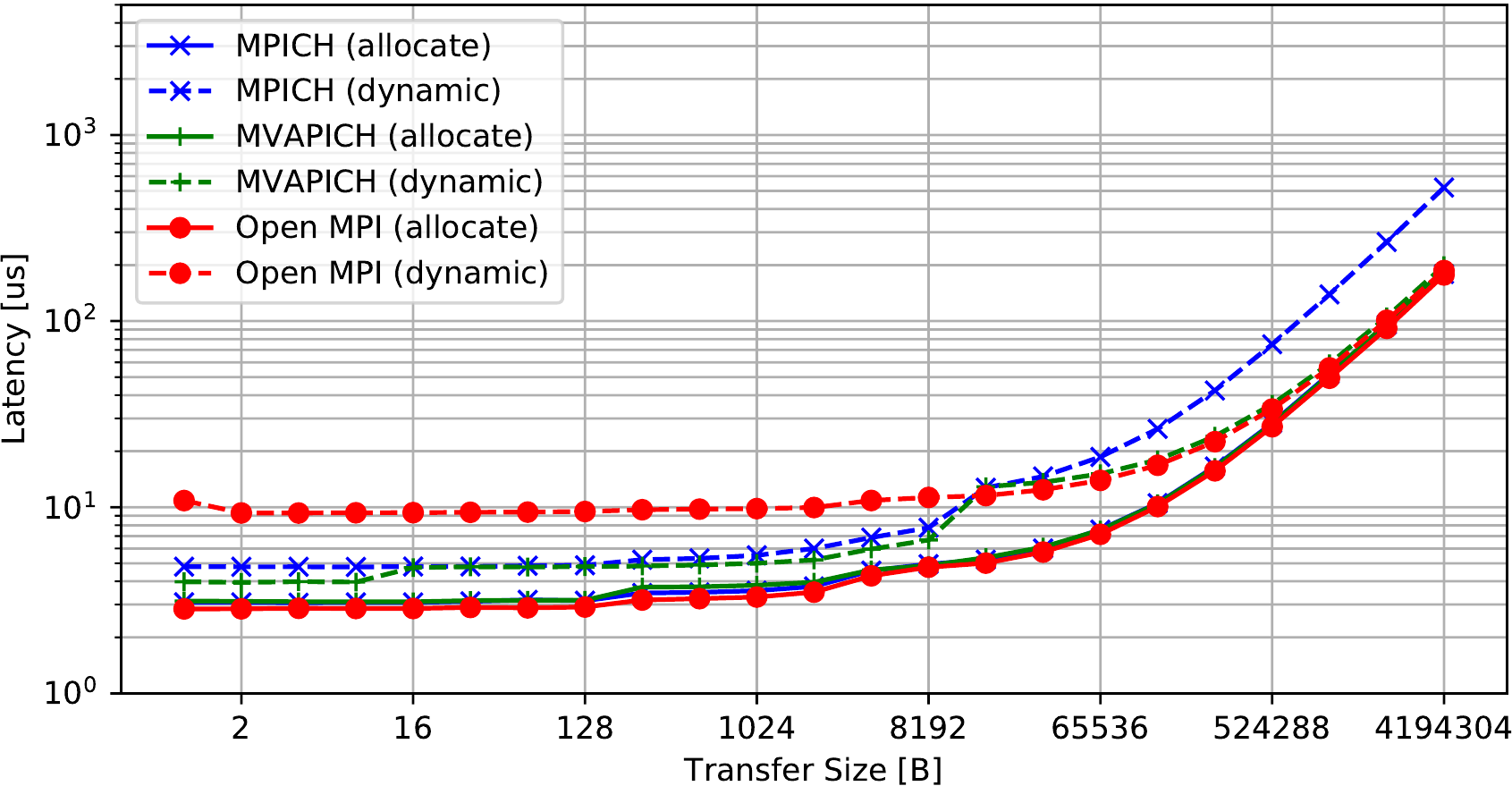}
\caption{Latency of put operations using allocated and dynamic windows.}
\label{fig:osu_latency}
\end{figure}

\autoref{fig:osu_latency} shows the latency of \code{put} operations measured using the OSU benchmark \code{osu\_put\_latency} benchmark on different MPI implementations.
While the latencies on allocated windows are similar across the three implementations, the differences are significant for dynamic windows.
Especially for smaller transfer sizes, the penalty of using dynamic windows over allocated windows ranges from a factor of 1.5$\times$ (MPICH, MVAPICH) over 3$\times$ (Open MPI using UCX).
As described earlier, this discrepancy between allocated and dynamic windows stems from the missing registration information, which either leads to a fallback to AM-based emulation or requires fetching the registration information before issuing the actual operation.

\subsubsection{One-Sided Progress Behavior}

\begin{figure}
\includegraphics[width=.9\columnwidth, page=16]{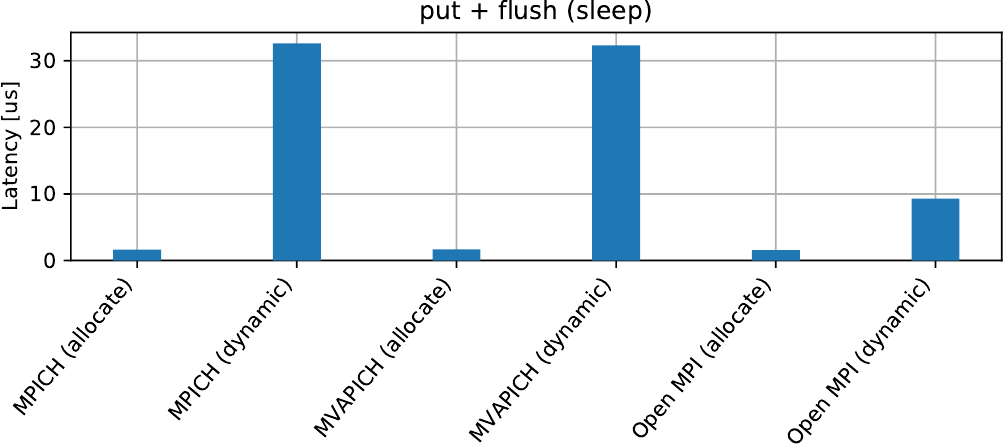}
\caption{Average latency of 100,000 single-byte put and flush with the target sleeping for 3\,s. Latencies above $30\,\mu s$ indicate no progress while the target process is not executing MPI calls.}
\label{fig:one_sided_progress}
\end{figure}

In order to better understand the behavior of the implementations, we repeat a benchmark here that was used in~\cite{Schuchart:2019:UMR} to determine the one-sided behavior of various accumulate MPI implementations.
In this test, the target process is busy outside of MPI for a fixed amount of time before waiting in an MPI barrier for the origin to complete the execution of a number of RMA operations.
For the results shown in \autoref{fig:one_sided_progress}, the origin performs $n = 100\,000$ put operations, each followed by a flush, while the target is busy outside of MPI for $t = 3\,s$.
Thus, a latency of $\frac{t}{n} > 30\,\mu s$ indicates that the origin is not progressing until the target enters the MPI barrier.
As can be seen in \autoref{fig:one_sided_progress}, both MPICH and MVAPICH lack progress for dynamic windows, indicating an implementation relying on the target CPU to execute active messages (\autoref{fig:dynamic_win_put:am}).
Operations on dynamic windows using the UCX integration in Open MPI, on the other hand, progress, albeit at a significantly higher latency, as discussed in the previous section.
We note that while AM-based emulation may yield sufficiently low-latencies in benchmarks such as the \code{osu\_get\_latency}, in practice it renders the performance of MPI RMA unpredictable as performance depends on the behavior of the target process, (partially) defeating the purpose of a one-sided programming interface.

\subsection{Life-Time Control Through Memory Handles}

Allowing users to provide MPI directly with registration information on memory to be accessed through windows provides both life-time information and avoid additional overhead during communication operations.
We propose the following three additions to the MPI RMA interface (their signatures are detailed in \autoref{lst:mpi_mem_handle}):

\begin{description}
\item[\code{MPIX\_Memhandle\_create}] registers a memory region starting at \code{base} of size \code{size} with the provided dynamic window for later access through RMA operations. The function returns in \code{memhandle} a memory handle of size \code{memhandle\_size}. The \code{memhandle} should be a buffer of at least \code{MPI\_MAX\_MEMHAN\-DLE\_\-SIZE} bytes. The memory handle contained in this buffer can be distributed to peer processes.
\item[\code{MPIX\_Win\_from\_memhandle}] The received memory handle is passed to this function together with the same dynamic window. The function returns a new window object whose only allowed target is the provided \code{target} and the usual configuration of displacement unit, size, and info to control aspects of the newly created window.
\item[\code{MPIX\_Memhandle\_release}] Once all RMA operations have completed and the registered memory is not needed anymore (i.e., all peers have signaled completion) the memory handle can be released using this function. After a call to this function, no more RMA operations may be issued on windows created from this memory handle. The corresponding windows must be freed through a call to \code{MPI\_Win\_free}.
\end{description}

Instead of sending the virtual address of an attached memory region to the peer the application now sends the registration information directly, which is an opaque data structure that is specific to the underlying implementation (which may differ from platform to platform for the same MPI implementation).

The call to \code{MPIX\_Win\_from\_memhandle} is a local operation and the resulting window remains connected to its parent window.
We restrict synchronization of such windows to passive target synchronization and require the lock and unlock to be applied on the parent dynamic window.
These restrictions allow the MPI implementation to avoid allocating additional internal memory during the creation of the memory handle required to handle these synchronization operations.
Thus, the only operations permitted on memory handle windows are put, get, and accumulate operations as well as flushes.
We expect users to use shared locks and rely on other synchronization and signaling mechanisms such as collective operations, point-to-point operations, or accumulate operations.
By allocating a separate window object, the implementation is not required to maintain and repeatedly traverse a list of attached memory handles but instead the resulting window identifies the remote memory region directly and allows for the implementation to stop tracking that remote memory region once the applications calls \code{MPI\_Win\_free} on the memory handle window.

\begin{listing}
\begin{lstlisting}
/* Maximum size of a memory handle, implementation specific */
#define MPI_MAX_MEMORY_HANDLE_SIZE <value>

/* Start exposure for memory and return handle to be sent
 * to peers. The memory handle is returned in memhandle
 * and has the actual size memhandle_size. The memhandle
 * argument should be a byte array of at least
 * MPI_MAX_MEMORY_HANDLE_SIZE elements. */
int MPIX_Memhandle_create(
  void *base, MPI_Aint size, 
  MPI_Info info, MPI_Win parentwin,
  void *memhandle, int *memhandle_size);

/* Create a window from a memory handle. The data
 * in memhandle should have been filled in by a call to
 * MPIX_Memhandle_create and sent to a peer or used
 * locally. The parentwin argument is a previously allocated
 * dynamic window. A newly created window will be returned
 * in newwin. */
int MPIX_Win_from_memhandle(
  const void *memhandle, 
  MPI_Aint size, int disp_unit,
  MPI_Info info, int target, 
  MPI_Win parentwin, MPI_Win *newwin);

/* Release a memory handle, ending the associated memory's
 * exposure. The data in memhandle must have previously
 * been filled in by a call to MPIX_Memhandle_create.
 * It is erroneous to release a memory more than once. */
int MPIX_Memhandle_release(void *memhandle, MPI_Win parentwin);
\end{lstlisting}
\caption{The MPI Memory Handle interface.}
\label{lst:mpi_mem_handle}
\end{listing}

\subsection{Example}

\begin{figure}
\includegraphics[width=\columnwidth]{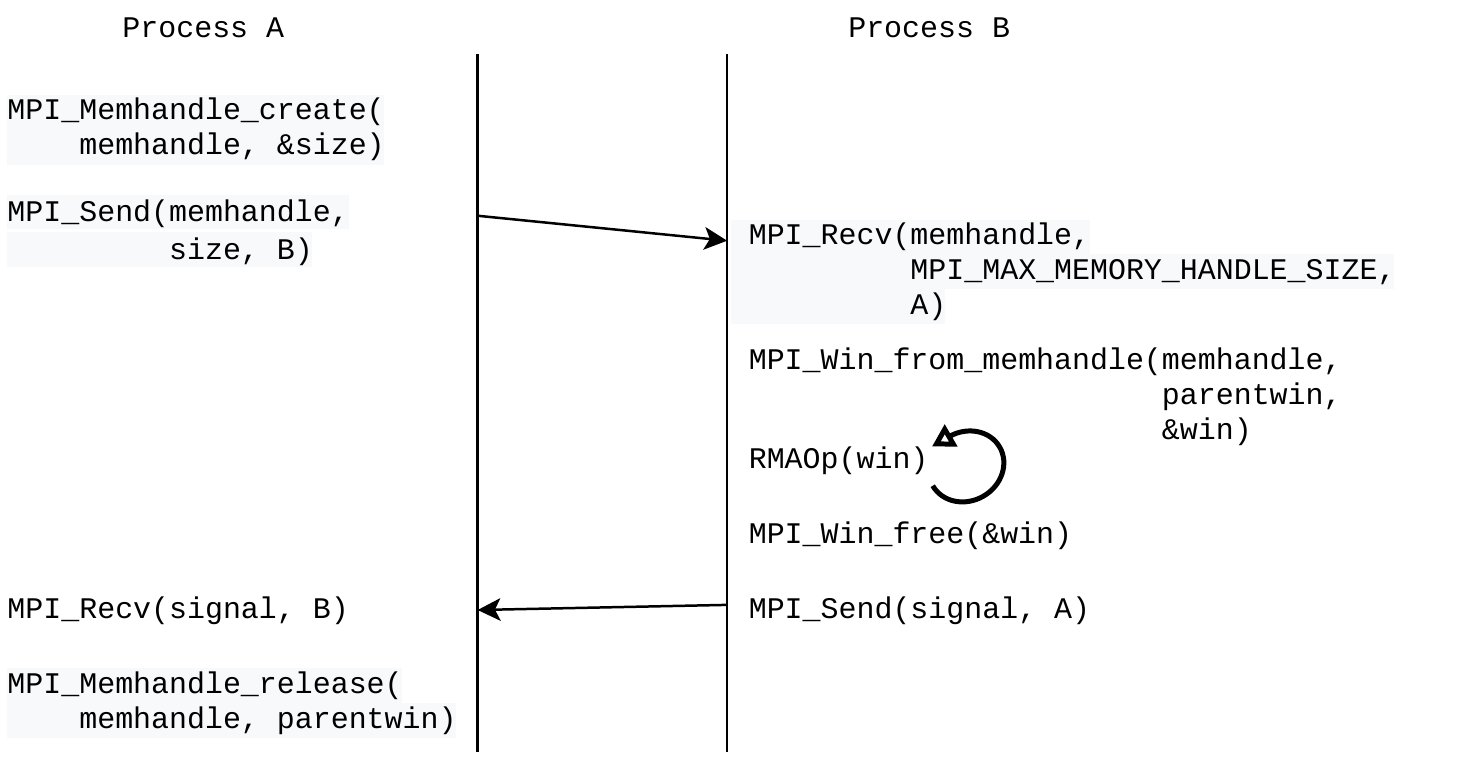}
\caption{An example of using memory handles and the associated memory handle windows. \code{RMAOp} signifies any RMA operation on the window.}
\label{fig:memhandle_usage}
\end{figure}

An example for how we envision memory handles to be used is provided in \autoref{fig:memhandle_usage}.
After creating the window \code{win} from the memory handle received from Process A, Process B performs an arbitrary number of RMA operations on the corresponding target memory region before signaling completion back to Process A, which then releases the memory handle.
We will show in \autoref{sec:evaluation_memhandles} that latencies using memory handle windows are on par with allocated windows in our proof-of-concept implementation.

\section{Further Improvements}
\label{sec:additional_improvements}

In this section we briefly discuss efforts that we consider beneficial for the future direction of the MPI RMA API but that are beyond the scope of this paper.

\subsection{Completion Notification}
\label{sec:notify}

Past work has focused on completion notification at the target and we support the approach presented in~\cite{Sergent:2019:ENM}, which was based on a previous proposal~\cite{Belli:2015:Notified}.
However, we caution that instead of introducing new test/wait routines (\code{MPIX\_Win\_test\_\-notify} and \code{MPIX\_Win\_\-wait\_no\-tify}) the notification should integrate with the existing request facilities.
Due to the nature of progress in MPI, these functions have to ensure progress in order to drive non-RMA communication operations on whose completion at the origin the notification may depend, e.g., collectives or point-to-point operations.

Using a request-based notification mechanism (without relying persistent requests~\cite{Belli:2015:Notified}) allows users to test or wait for completion all MPI-related communication, including RMA notifications.
We envision a function such as \code{MPI\_\-Win\_\-wait\_notify(win, notify\_id, request)} that returns a request, which can later be used with the regular request test and wait infrastructure in MPI.
We leave an in-depth investigation into such an interface for future work.

\subsection{Remote-Completing Request-Based Operations}
\label{sec:remote_complete}

MPI RMA provides variants of put, get, accumulate, and get-accu\-mulate that return a request that can be used to test and wait for completion of that particular operation.
However, the completion of a request returned by \code{MPI\_Rput}, \code{MPI\_Raccumulate}, and \code{MPI\_Rget\_accumulate} only signals local completion, requiring a subsequent flush to achieve remote completion before signaling completion to the target.
This flush, in turn, may wait for the completion of unrelated operations, potentially from other threads.
In many cases, MPI implementations are able to implement request-based put with remote completion more efficiently than the application using a flush (e.g., by leveraging guarantees of the underlying transport library). %, it can likely do so more efficiently than the application issuing a flush.
We thus propose adding functions \code{MPI\_Rrput} (\textbf{r}emote-completing \textbf{r}equest-base \textbf{put}), \code{MPI\_Rraccumulate}, and \code{MPI\_Rrget\_\-accumulate} with similar signatures as their current request-based counter-parts..
The completion of requests provided by these procedures signal both local and remote completion, alleviating the need for an additional flush and thus potentially improving the efficiency of applications using request-based operations.
An example of the case described above using a remote-completing request-based put is given in \autoref{lst:put_allreduce}.
If a regular call to \code{MPI\_Rput} was used instead, a flush would be necessary in Line~\ref{line:put_allreduce:1} before the call to \code{MPI\_Allreduce} was used to signal completion to all peers.

\begin{listing}
\begin{lstlisting}[numbers=left]
int flag, one = 1;
MPI_Request req;
MPI_Rrput(..., target, win, &req);
do {
  do_useful_work();
  MPI_Test(&req, &flag, MPI_STATUS_IGNORE);
} while (!flag);
/* A call to MPI_Win_flush would be required if MPI_Rput was used */ |\label{line:put_allreduce:1}|
MPI_Allreduce(&another_variable, ...);
\end{lstlisting}
\caption{The example of \autoref{lst:put_inc} using a remote-completing request-based put and a collective operation for synchronization.}
\label{lst:put_allreduce}
\end{listing}

\subsection{Deprecating Active Target Synchronization}

We encourage efforts to engage with users of the active target synchronization interface to identify potential road-blocks in the transition to passive target synchronization, with the goal of phasing out active target synchronization.
By eventually removing active target synchronization, the RMA part of the standard would become cleaner and more concise, removing a significant portion of its complexity and providing easier access to the one-sided communication chapter.
Moreover, we believe that the collective nature of active synchronization incurs excessive overhead and techniques mentioned earlier in this section may achieve similar goals more efficiently.
However, such an effort has to be undertaken in collaboration with the community, which is beyond the scope of this paper.
Previous surveys of MPI usage might help in identifying the relevant user groups~\cite{Laguna:2019:ALSS, Berholdt:2018:MPIUsage}.

\section{Implementation}
\label{sec:implementation}

\subsection{Reference Implementation}
\label{sec:ref_impl}

\begin{figure}
\centering
\includegraphics[width=.7\columnwidth]{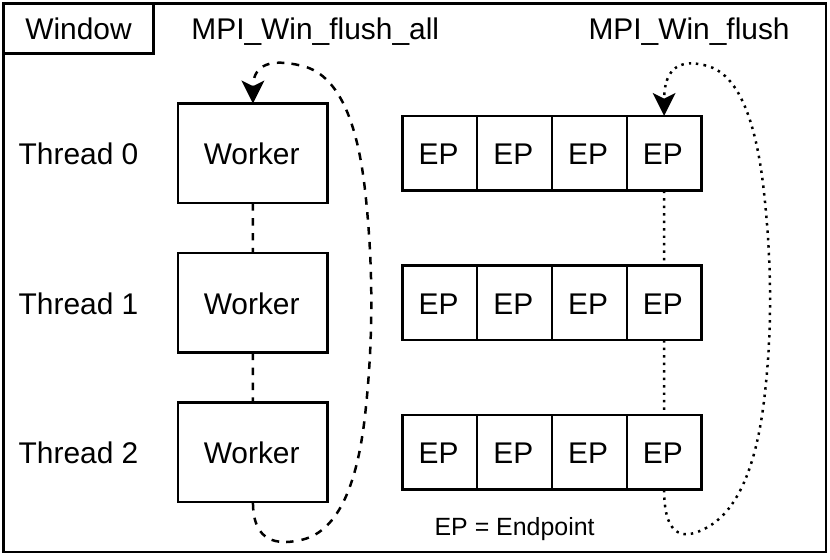}
\caption{Flushes in the UCX one-sided communication module in Open MPI. Threads calling \code{MPI\_Win\_flush} iterate over the endpoints of all threads to ensure completion of all operations issued by the process.}
\label{fig:osc_ucx}
\end{figure}

The used reference implementation (the UCX one-sided communication module in Open MPI's main development branch) employs thread-specific UCX worker objects for each window.
Since UCX endpoints are specific to a worker, each thread also manages its own set of endpoints (connections with peers in the window) that are created upon the first access to that peer in the window and used to issue RMA operations.
The worker and endpoints are stored in lists in the window, over which threads iterate during flushes on that window, as depicted in \autoref{fig:osc_ucx}.
Access to that list and to each thread's connection information are protected through mutexes to ensure thread-safety.
While this scheme allows threads to issue operations on independent endpoints, it leads to significant synchronization overheads during a flush operation.

Our proof-of-concept implementations are based on this infrastructure and we note that other implementations may have different approaches, leading to different degrees of effectiveness of the proposed RMA extensions.

\subsection{Thread-Scope Flushes}

The aforementioned reference implementation allowed for an easy implementation of thread-scope flushes discussed in \autoref{sec:thread_local_flush}: instead of iterating over the list of workers or endpoints, a thread calling into a flush only operates on its local worker or endpoint.\footnote{The proof-of-concept implementation for the thread-scope and operation ordering info keys can be found at \url{https://github.com/devreal/ompi/tree/mpi-win-dup-with-info}.}
Thus, no access to workers or endpoints owned by other threads is necessary, greatly reducing the amount of both the amount of work and inter-thread synchronization required to achieve completion of operations issued by an individual thread.

\subsection{Operation Ordering}
\label{sec:oporder_impl}

When issuing RMA operations on windows for which the \code{mpi\_win\_\-or\-der} info key discussed in \autoref{sec:oporder} is set to \code{true}, the calling thread calls into \code{ucp\_worker\_fence} on the used UCX worker before issuing the actual operation.
This function call ensures that the operation will complete at the target only after all previously issued operations have completed at the target.

The used UCX worker depends on the \emph{scope} set for the window.
Since \code{ucp\_worker\_fence} guarantees operation ordering for a specific worker only, all operations are funneled through the endpoints of a single worker when the \emph{process scope} is enabled on the window (the default).
While this may incur additional synchronization between threads, it allows for operation ordering without explicitly waiting for operations to complete using a full flush.
With \emph{thread scope} enabled, the calling thread invokes \code{ucp\_worker\_fence} on its local worker.

\subsection{Hardware Accumulate Operations}

Open MPI already provides support for an info key on windows, called \code{acc\_single\_intrinsic}, that allows users to signal to the implementation that only single-element accumulate operations with support for intrinsic operations will be used.
It's effectiveness for ensuring low-latency accumulate operation on supported hardware has been shown in~\cite{Schuchart:2019:UMR}.
Due to space limitations, we refrain from any further discussion of both the implementation and its effectiveness as the proposal presented in \autoref{sec:hardware_accumulate} provides the same guarantees as the existing \code{acc\_single\_intrinsic} info key.

\subsection{Memory Handle Windows}

In contrast to collectively allocated windows, memory handle windows only provide access to a single memory region at a specific target process.
It is thus sufficient to store the information for that target in the window.\footnote{The proof-of-concept implementation for memory handle windows can be found at \url{https://github.com/devreal/ompi/tree/osc-win-memhandle-parentwin}.}
Our implementation employs the parent window UCX worker and endpoint information described in \autoref{sec:ref_impl} for communication and only stores the memory handle's registration information in the window, allowing for fast creation (and destruction) of memory handle windows.

\section{Evaluation}
\label{sec:evaluation}

\subsection{Thread-Scope Flushes}

We use the RMA-MT benchmark to measure latencies of RMA operations in a multi-threaded context~\cite{Dosanjh:2016:RMAMT}.
The existing benchmark covers both active and passive target synchronization, with multiple worker threads issuing RMA operations and the main thread performing the ensuing RMA synchronization.
While this pattern may be useful for fork-join thread models such as OpenMP work-sharing loops, it is inadequate for task-based applications that typically do not exhibit synchronization points.
We have thus extended the benchmark to include a variant in which threads perform RMA operations followed by flushes, a pattern that is commonly found in applications using puts and flushes to ensure remote completion.

\begin{figure}
\begin{subfigure}{.8\columnwidth}
\includegraphics[width=\textwidth, page=1]{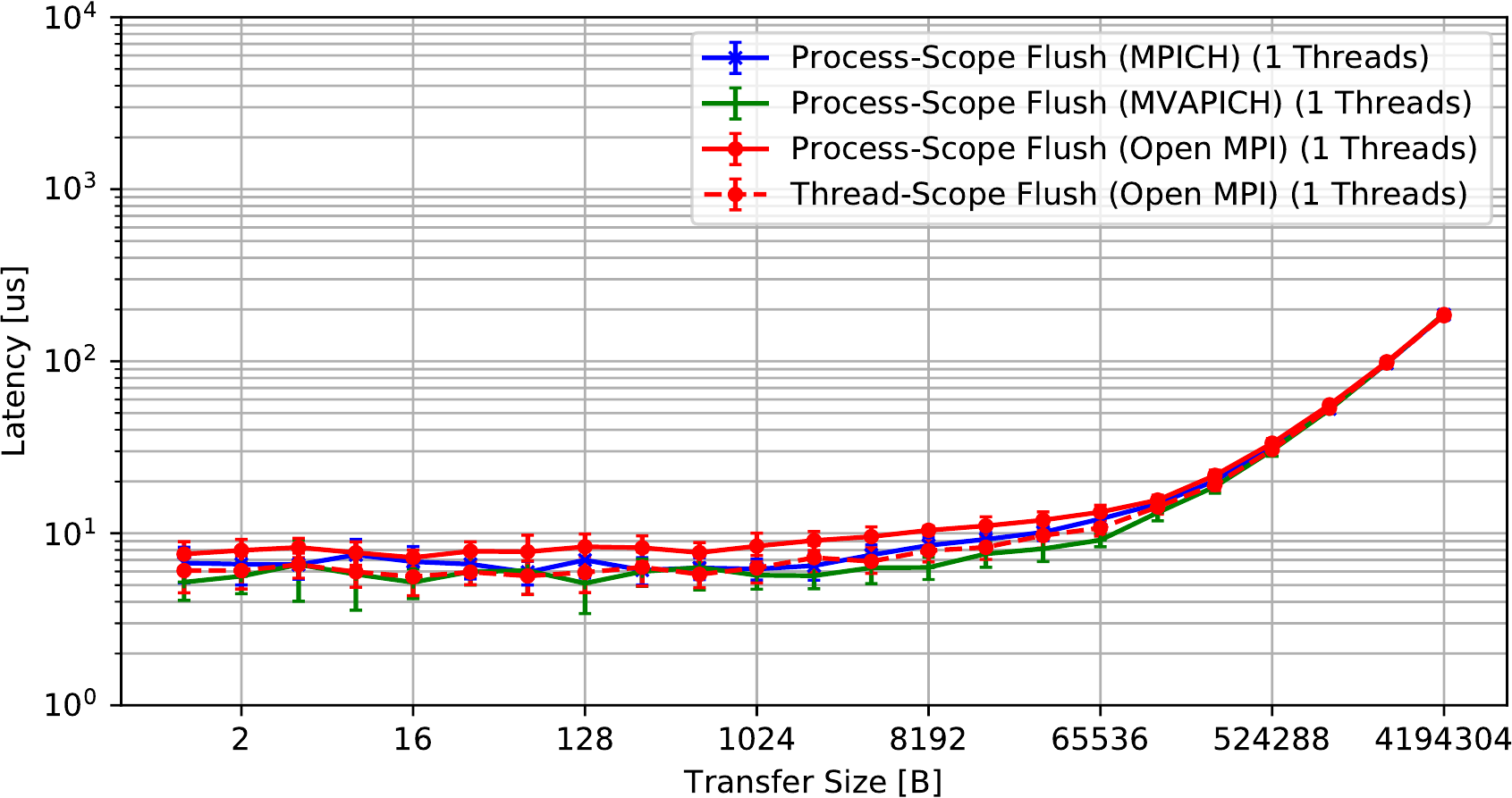}
\caption{Single Worker Thread}
\label{fig:mtflush_sizescale:1}
\end{subfigure}

\begin{subfigure}{.8\columnwidth}
\includegraphics[width=\textwidth, page=6]{figures/rma-mt-putflush_1199727.hawk-pbs5-crop.pdf}
\caption{32 Worker Threads}
\label{fig:mtflush_sizescale:32}
\end{subfigure}
\caption{Latency of multi-threaded put and flush with process-scope and thread-scope flushes.}
\label{fig:mtflush_sizescale}
\end{figure}

The latencies of a put followed by a flush when selecting either thread- or process-scope using the \code{mpi\_win\_scope} info key (\autoref{sec:thread_local_flush}) with Open MPI as well as using MPICH and MVAPICH are shown in \autoref{fig:mtflush_sizescale}.
The slightly higher latency of process-scope flushes in the case of a single worker thread shown in \autoref{fig:mtflush_sizescale:1} for Open MPI can be explained by the fact that the single worker thread has to perform a flush on its endpoint and the main thread's endpoint, which is not required when using thread-scope flushes.
Overall, however, the latencies of the different implementations are mostly similar.
By contrast, for 32 worker threads shown in \autoref{fig:mtflush_sizescale:32}, the use of thread-scope flushes leads to an order of magnitude lower latencies for small transfer sizes compared to MPICH and MVAPICH and close to two orders of magnitude for Open MPI. 
For larger transfer sizes a factor of two is achieved.

\begin{figure}
\centering
\includegraphics[width=.8\columnwidth, page=7]{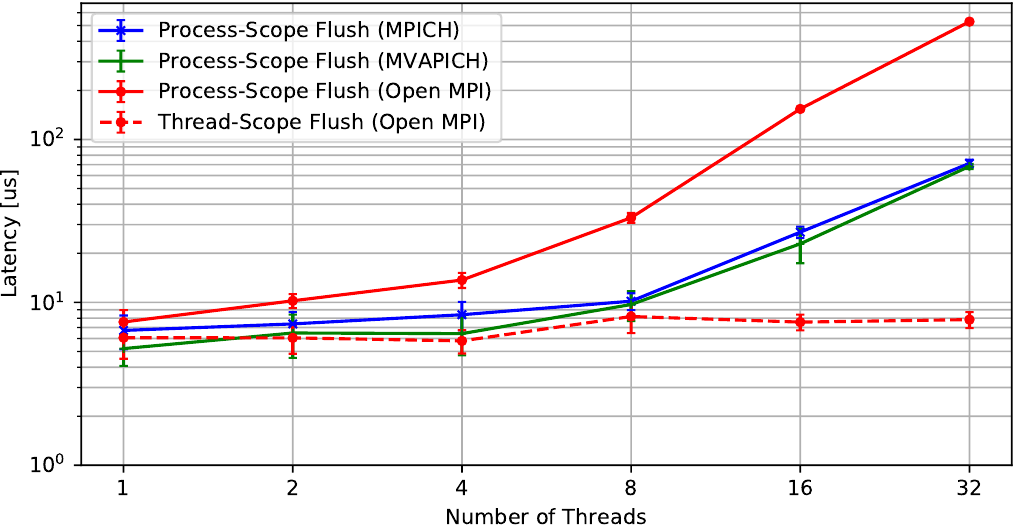}
\caption{Latency of single-byte put and flush with process-scope and thread-scope when scaling the number of threads.}
\label{fig:mtflush_threadscale}
\end{figure}

The thread-scaling behavior for single-byte transfer sizes is shown in \autoref{fig:mtflush_threadscale}.
The benefit of using thread-scope flushes (where appropriate) becomes clear, as it reduces the amount of work each thread has to perform inside a flush, reducing the inter-thread synchronization to a minimum.

\subsection{Operation Ordering}

\iffalse
\begin{figure}
\begin{subfigure}{.8\columnwidth}
\includegraphics[width=\textwidth, page=1]{figures/rma-mt-putflush_order_1199501.hawk-pbs5-crop.pdf}
\caption{Single Worker Thread, Process-Scope}
\label{fig:oporder_lat:1}
\end{subfigure}

\begin{subfigure}{.8\columnwidth}
\includegraphics[width=\textwidth, page=6]{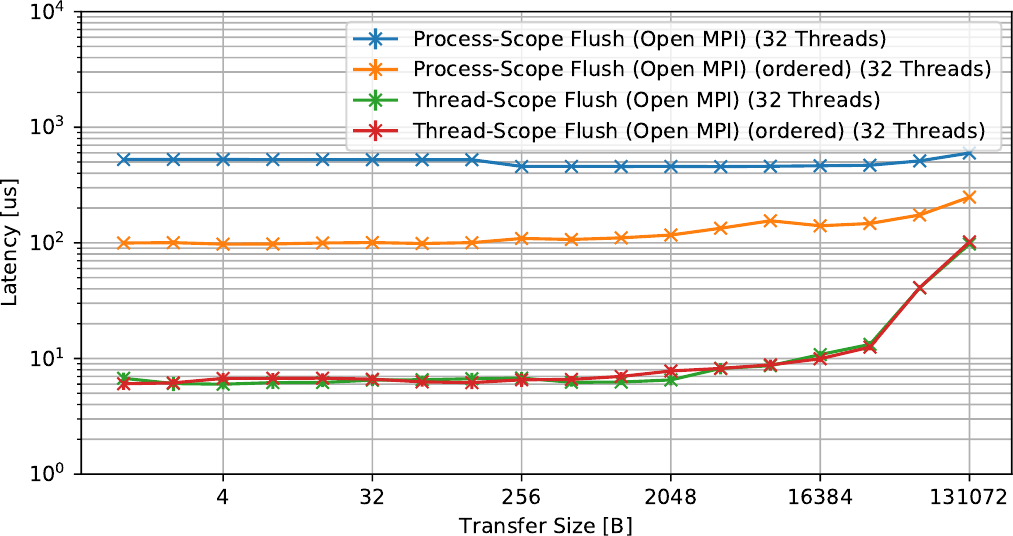}
\caption{32 Worker Thread, Process- and Thread-Scope}
\label{fig:oporder_lat:32}
\end{subfigure}
\caption{Latencies of put and flush with and without operation ordering enabled.}
\label{fig:oporder_lat}
\end{figure}
\fi

\begin{figure}
\centering
\includegraphics[width=.8\columnwidth]{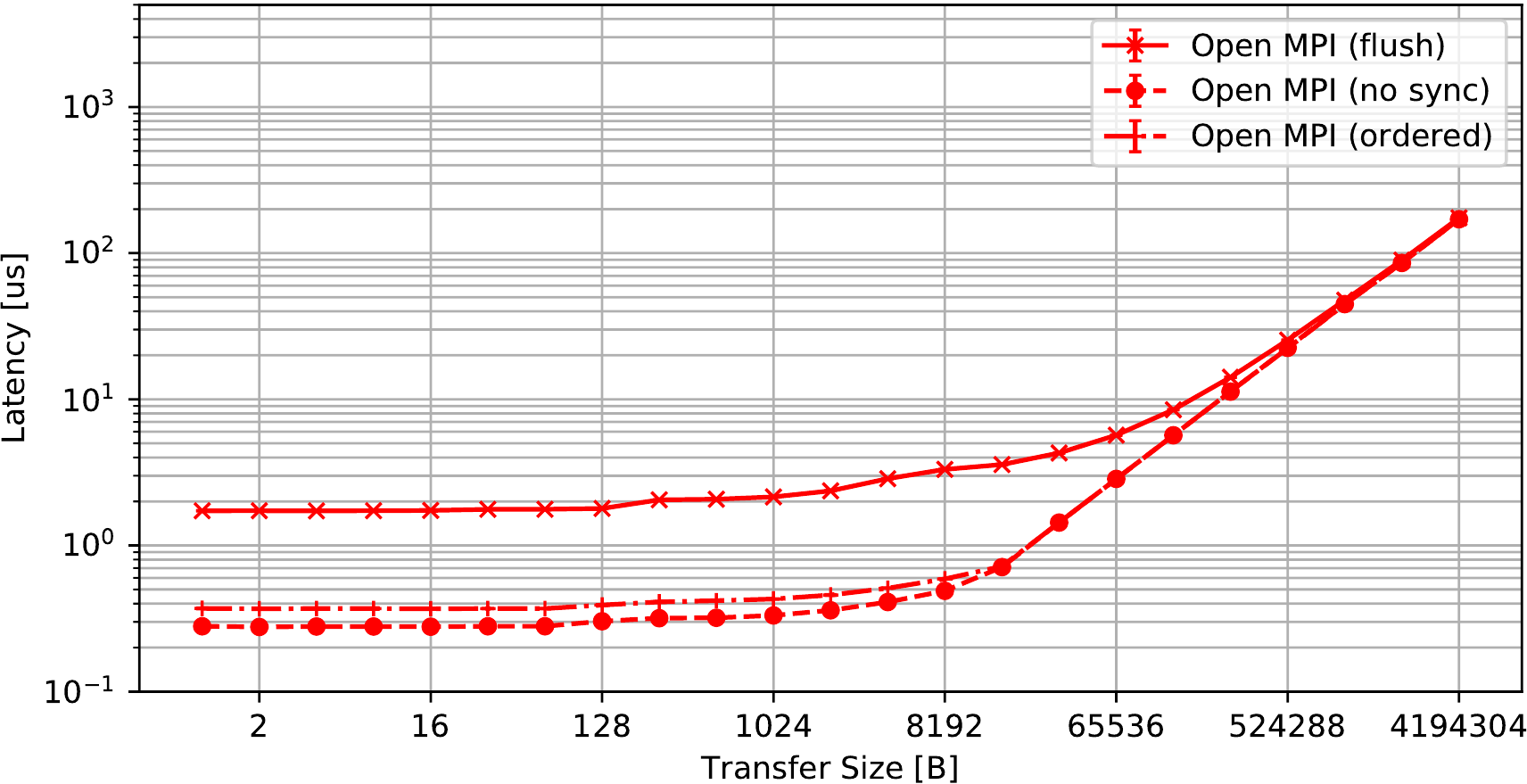}
\caption{Latency of put with flush as well as put without intermediate synchronization but ordering enabled using the \code{mpi\_win\_order} info key.}
\label{fig:oporder_lat}
\end{figure}

\begin{figure}
\includegraphics[width=.8\columnwidth, page=6]{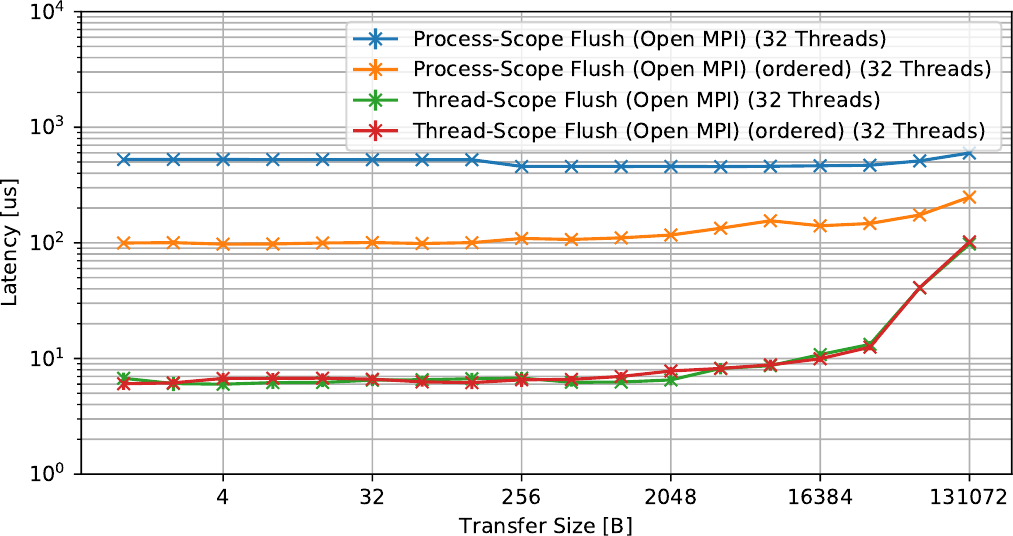}
\caption{Latencies of put and flush using 32 worker threads with and without operation ordering enabled.}
\label{fig:oporder_lat_mt}
\end{figure}

We have modified the \code{osu\_put\_latency} benchmark to include an option suppressing intermediate flushes, i.e., puts are issued in a loop and synchronization happens during \code{MPI\_Win\_unlock} after the specified number of operations have been started.
This allows us to better observe the overhead of operation ordering.
\autoref{fig:oporder_lat} shows the latency for regular put and flush as well as the variant without intermediate synchronization, both with and without operation ordering enabled using the \code{mpi\_win\_order} info key discussed in \autoref{sec:oporder}.
While some additional latency can be observed due to the requested ordering, the latency of ordered puts is still significantly lower than the if flushes were used to enforce ordering of RMA operations.
While this is by no means surprising (flushes likely incur at least the latency of a full network round trip), it underscores that the MPI RMA interface should provide means for ordering operations beyond waiting for completion.

We use the same RMA-MT benchmark to compare the impact of operation ordering using the \code{mpi\_win\_order} info key discussed in \autoref{sec:oporder} on the latency of put operations.
With 32 worker threads (\autoref{fig:oporder_lat_mt}), enabling operation ordering with process-scope flushes reduces latencies since, as described in \autoref{sec:oporder_impl}, all operations are funneled through a single endpoint. %, reducing the work to be performed inside the flush to a single endpoint, rather than 32 endpoints to be flushed by each thread.

\subsection{Memory Handles}
\label{sec:evaluation_memhandles}

\begin{figure}
\centering
\includegraphics[width=.8\columnwidth, page=1]{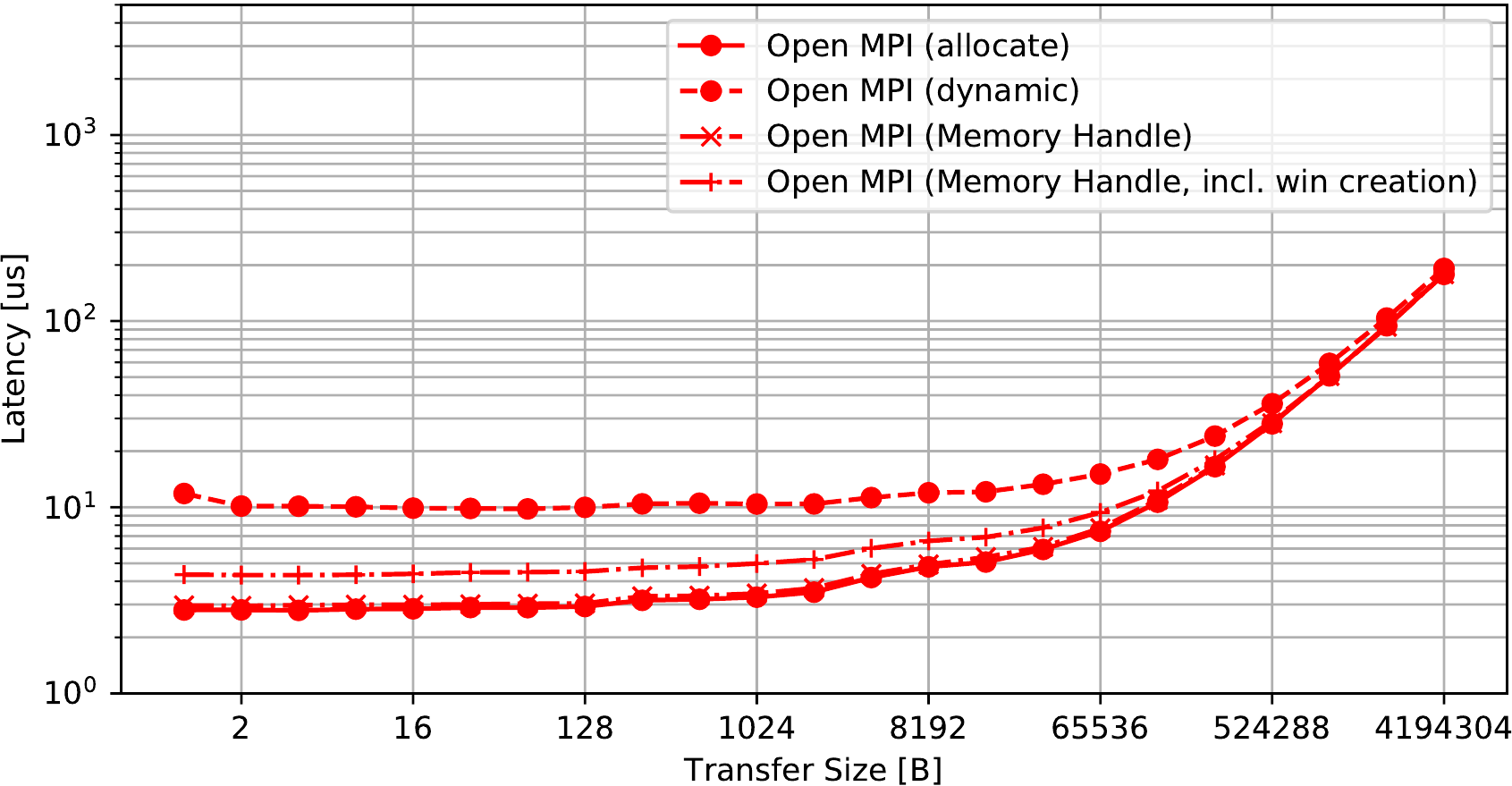}
\caption{Latency of put using allocated windows and memory handle windows.}
\label{fig:memhandle_lat}
\end{figure}

Similar to the previous results, the latency of puts measured using the \code{osu\_put\_latency} using allocated, dynamic, and memory handle windows are shown in \autoref{fig:memhandle_lat}.
The difference between allocated windows and windows created from memory handles is negligible.
Compared to the latencies of today's dynamic windows discussed in \autoref{sec:dmh}, the benefits of combining the flexibility of dynamic windows with the use of direct RDMA without the additional overhead of querying registration information become clear.

Latencies with added window creation and destruction are included in \autoref{fig:memhandle_lat}, adding approximately $1\,\mu s$ and still being significantly lower than the latencies for existing dynamic windows while employing the network's RDMA capabilities.
We believe that such an overhead is sufficiently low to allow applications to rapidly create and destroy memory handle windows for use with RMA operations. 

\iffalse
\begin{figure}
\includegraphics[width=.9\columnwidth, page=2]{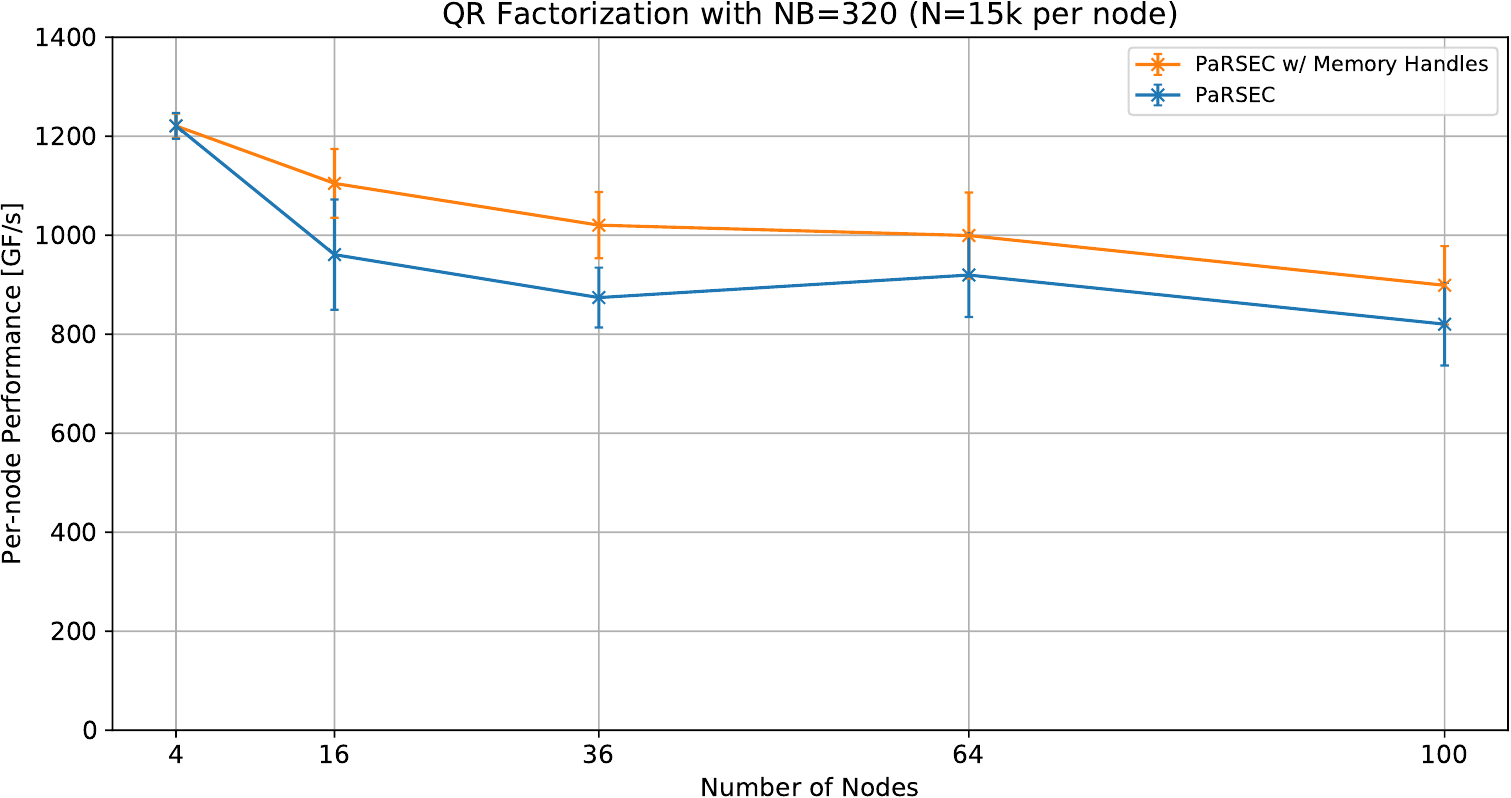}
\caption{DPLASMA QR factorization with PaRSEC using memory handle windows ($N = 15k$ per node, $NB=320$).}
\label{fig:qr_memhandle}
\end{figure}

In PaRSEC's remodeled (still experimental) communication engine, active messages are used to exchange activation messages for tasks and the necessary input data is communicated using a one-sided communication abstraction.
Because of the limitations of MPI RMA described earlier and the highly dynamic nature of the data flow in PaRSEC, the latter is currently implemented using internal active messages and two-sided MPI communication.
We have thus integrated memory handle windows to replace the emulated one-sided communication with put operations on memory handle windows.
\autoref{fig:qr_memhandle} compares the performance of QR factorization implemented in DPLASMA (which runs on top of PaRSEC~\cite{DAGuE:2011, Bosilca:2011:FDD}) using emulated one-sided communicaton against the use of memory handle windows.
In both variants, synchronization happens through active messages as the limited MPI tag space may not be sufficiently large to uniquely identify all data flows.
Using memory handle windows, however, the number of requests to be managed is essentially cut in half as no explicit receives for data are required.
The number of active messages is the same in both variants.

\fi

\section{Related Work}
\label{sec:related_world}

Several improvements to MPI's ability of handling multi-threaded communication has been proposed over the years, ranging from thread-safe probes~\cite{Hoefler:2010:EMS} over thread-specific endpoints~\cite{Dinan:2013:EMI} to partitioned communication~\cite{Grant:2019:FPM} and (most recently) the use of continuations~\cite{Protze:2020:MPID, Schuchart:2020:Fibers}.
Work on improved implementation support for multi-threaded MPI in general~\cite{Patinyasakdikul:2019:GMT} and RMA in particular~\cite{Hjelm:2018:IMM} has also been described.
The thread-scope flushes proposed in this work provide additional information to the implementation to better leverage some of that earlier work.

Several abstractions for one-sided communication provide collective allocation capabilities, including OpenSHMEM~\cite{Chapman:2010:OSHM, OSHMEM:1.5:2020} and GASNet~\cite{Bonachea:2018:GNE}, but lack local allocation of exposed memory.
Lower-level PGAS abstractions such as GASPI~\cite{Alrutz2013} and LCI~\cite{Dang:2018:LCI} provide dynamic local allocation of exposed memory.
The memory handle windows proposed in this work aim at closing the gap to these low-level abstractions and increase the flexibility of MPI RMA.

OpenSHMEM has introduced so-called contexts to provide isolation between threads, at the cost of significant extension of the API~\cite{Dinan:2014:CMH}.
The proposed duplicate windows with thread-scope setting is an attempt to achieve a similar to goal, without significantly extending the RMA interface.

The proposed memory handle windows may be useful for applications to work around limitations of hardware tag matching engines~\cite{Levy:2019:ETB} by reducing the number of exchanged messages, e.g., by organizing multiple data transfers through MPI RMA and using matched messages for signaling purposes only.
The proposed info key for ordering RMA operations might enable implementations to utilize triggered operations on network hardware, which already have proven useful in the implementation of collective operations~\cite{Islam:2019:MUH} and in the implementation of fence operations in OpenSHMEM~\cite{Flajslik:2015:Fence}.

\section{Conclusions}
\label{sec:conclusions}

We have identified several short-comings of the RMA part of the current MPI standard versions that potentially cause low performance due to high costs of synchronization and a lack of usage of available hardware resources in RMA operations.
By allowing users to provide additional information on the anticipated usage of windows, implementations can adapt to the application's usage patterns, enabling improved performance, e.g., by constraining the scope of flushes, reducing the number of flushes by enforcing the ordering of operations, and by constraining the number of elements in accumulate operations.
By introducing the duplication of windows, we provide a way for users to maintain differently configured handles to the same window resources, facilitating easy switching between configurations in different parts of an application.
Additionally, we propose to add the notion of memory handles to the MPI RMA interface, enabling bare-metal performance of dynamic windows by allowing users to manage memory registration information and provide life-time guarantees of memory segments, which eliminates costly querying at the target before performing RMA operations.
Our benchmarks show that the proposed additions to the RMA chapter can greatly reduce the synchronization overhead, allowing applications to make better use of the hardware capabilities through the MPI RMA interface.

\begin{acks}
The research leading to these results has received funding from the European Union’s Horizon 2020 research and innovation programme under the ChEESE project, grant agreement No. 823844.
This material is based upon work supported by the National Science Foundation under Grant No. \#1664142 and the Exascale Computing Project (17-SC-20-SC), a collaborative effort of the US Department of Energy Office of Science and the National Nuclear Security Administration.

\end{acks}

\bibliographystyle{ACM-Reference-Format}
\bibliography{references}

\end{document}